\documentclass[10pt, twoside]{article}
\usepackage{fullpage}
\usepackage{graphicx}
\usepackage{amssymb}
\usepackage{amsmath}
\usepackage{booktabs}
\usepackage{array}
\usepackage{multirow}
\allowdisplaybreaks
\usepackage{url}

\usepackage[authoryear,round,comma,compress]{natbib}
\bibpunct{(}{)}{,}{a}{}{,}

\title{End-to-end Planning of Fixed Millimeter-Wave Networks}
\author{Guan Pang, Jing Huang, Manohar Paluri (Facebook Computer Vision AML),\\
Brian Karrer, Onur Filiz, Birce Tezel, Nicolas Stier-Moses (Facebook Core Data Science),\\
Vish Ponnampalam, Tim Danford (Facebook Connectivity Lab)}
\begin{document}
\maketitle

\section*{Abstract}

This article discusses a framework to support the design and end-to-end planning of fixed millimeter-wave networks. Compared to traditional techniques, the framework allows an organization to quickly plan a deployment in a cost-effective way.
We start by using LiDAR data---basically, a 3D point cloud captured from a city---to estimate potential sites to deploy antennas and whether there is line-of-sight between them. With that data on hand, we use combinatorial optimization techniques to determine the optimal set of locations and how they should communicate with each other, to satisfy engineering (e.g., latency, polarity), design (e.g., reliability) and financial (e.g., total cost of operation) constraints. The primary goal is to connect as many people as possible to the network.
Our methodology can be used for strategic planning when an organization is in the process of deciding whether to adopt a millimeter-wave technology or choosing between locations, or
for operational planning when conducting a detailed design of the actual network to be deployed in a selected location.

\section{Introduction}

This article describes a suite of tools to simplify the end-to-end planning cycle of fixed millimeter-wave backhaul and access networks. Our primary motivation for this work has been to support rapid and cost-effective deployments of Terragraph,
%This article describes a suite of tools put forward to simplify %the end-to-end
%planning cycle of Terragraph networks, 
a technology developed by Facebook to
replace fibers and create high-speed backhaul networks \citep{tg}.
Terragraph is a 60 GHz, multi-node wireless system focused on bringing high-speed internet connectivity to dense urban areas. Utilizing commercial off-the-shelf components, the Terragraph system is optimized for high throughput at low cost. 
%The benefits of using the 60 Ghz band is that it is unlicensed %in several countries, similarly to the Wi-Fi 2.4 GHz and 5 GHz
%bands.  Terragraph's wireless system consists of radios designed for consumer electronics that contribute to being relatively inexpensive compared to traditional telecom infrastructure .  
Millimeter-wave backhaul and access networks, similar in architecture and design to Terragraph, are expected to be a key component of 5th generation (5G) mobile communication systems \citep{pi}.

Planning deployments of millimeter-wave networks poses unique challenges, not encountered during roll-outs of previous generations of networks. First, the high path loss of millimeter-wave signals and densification of the corresponding access network leads to relatively large numbers of nodes per square km. Manual planning is extremely time consuming and inefficient, thus not scalable. Second, as diffraction is not significant at these frequency ranges, line of sight or at least a first-order reflected path is required to form a link. Hence accurate and deterministic knowledge of street-level clutter is required to predict wireless connectivity between nodes. Finally, the mesh topology of these networks requires network optimization techniques to maximize service availability and other key metrics. Incorporating such constraints when doing a manual design is extremely difficult, even for relatively small-scale networks and for an experienced network designer.

%Given the limited range of the 60 GHz
%signal, these nodes are placed across a city at 200-250 meter intervals. The
%vast bandwidth and unique signal-absorbing nature of the band limits
%interference and simplifies network planning, while the unlicensed nature of
%the spectrum helps to further minimize costs. Designed to provide street level
%coverage, Terragraph implements a phase array antenna to retain the highly
%directional signal required for 60 GHz, but makes it steerable to communicate
%over a wide area. Given the architecture of the network, Terragraph is able to
%route and steer around interference typically found in dense urban
%environments, such as tall buildings or internet congestion due to high user
%traffic.  Its reduced interference and ability to operate in non-line-of-sight
%conditions increases customer reach. For customers or business in
%multi-dwelling units or high-rises, the Terragraph system can be externally
%attached to a building and connected to an in-building Ethernet data network.
%Combined with Wi-Fi access points, Terragraph is one of the lowest cost
%solutions to achieve 100 percent street-level coverage of gigabit Wi-Fi.

In this article, we describe a fully-automated network planning system put forward to overcome the above-mentioned challenges. After selecting a potential area where Terragraph is going to be deployed, we capture 3D LiDAR ({\em Light Detection and Ranging}) images of the area. Using computer vision and machine learning techniques, we automatically identify potential sites where we can deploy network nodes. We also determine {\em line-of-sight} (LoS) path availability between those sites. Typically, only a fraction of these potential sites will be required in the final network but the more sites that are available as potential sites the more flexibility the designer has to optimize the key metrics. After digital location data is generated from the images, we augment our dataset with the location of the available {\em Fiber POPs} (point-of-presence) and demand information. Although we will not expand much on demand information in this article, it roughly consists on determining the bandwidth required to be fulfilled in a set of points in the map which is what determines how much supply must be built.
We use the network data generated from the images, the Fiber POPs and the demand information to build an integer programming model which is fed into a optimization solver that down-selects sites and designs an actual network.
The objective function is to minimize the {\em total cost of ownership} while satisfying a variety of constraints
that ensure that the network is adequate. The constraints include satisfying demand and respecting engineering design limits for latency, reliability, capacity, polarity, etc.

The remainder of this article provides details of our planning system and methodology. We start with a brief literature review in Section~\ref{litrev} to position our methodology with respect to other methods. 
Section~\ref{cv} describes how we process the LiDAR data to generate a list of potential sites and connections between those locations. 
In Section~\ref{ip}, we explain the integer programming formulation that is used to design a network consisting of a subset of sites, connections and POPs, and we offer concluding remarks in Section~\ref{concl}.

\section{Literature Overview}\label{litrev}

Given a set of nodes and links forming a graph, the network design problem aims to find a subgraph optimizing an objective function in which certain criteria are met. This problem arises in a broad range of practical applications such as logistics, transportation, supply chain and telecommunications. Since it is a classical problem there is a vast literature on this subject. We refer the reader to two of the earlier surveys on general network design problems by \citet{wong1976survey} and \citet{minoux1989networks} for a general background, and \citet{Kennington-book} for a book specifically about wireless network design. 
Furthermore, \cite{winter1987steiner} surveys existing heuristics and exact algorithms for Steiner problem, which is a version of a network design problem in which one wants to connect a set of access points (also known as leaves, see Section~\ref{steiner}). 
In the area of telecommunications, \cite{klincewicz1998hub} and \cite{campbell2012twenty} review the literature on hub location problems in backbone network design.
For network design in transportation, \citet{magnanti1984network} provide a great overview of models and algorithms. 
For network design in supply chains and facility location, \cite{shen2007integrated} and \cite{melo2009facility} discuss many interesting applications and models. 
In relation to wireless network design, \cite{pathak2011survey} and \cite{benyamina2012wireless} review the design of mesh networks. 
They discuss several crucial challenges such as finding the right topology, what routing protocol to use, as well as how to capture interference and reliability. 
Although our network design problem has similarities to these models, and to facility and hub location problems, it has many aspects that distinguish it from the design of mesh networks. 
Indeed, these models typically consist of a single-stage design problem, whereas millimeter-wave network design (including Terragraph) is a hierarchical problem:
they networks require (1) the selection of street furniture to set up sites, which typically require conditioning and rental, and (2) the placement of antennas (called nodes) on sites which typically require acquiring and installing the harding and then covering operational expenses such as electricity consumption. The definition of an edge in the network design problem requires to consider both layers in the hierarchy since communication between sites is possible only if antennas pointing to each other are deployed in two sites that are within each other's line of sight.

From the computer vision perspective, earlier work on road furniture detection, such as \cite{doubek2008mobile}, was based on images and videos. However, 3D LiDAR data have become more widespread because one can avoid the illumination and background confusion problems that frequently occur with 2D images. In terms of pole-like object detection from point clouds, some of the proposed slicing-based methods can only handle antenna attachments at the bottom or on the top but not in the middle of the trunks \citep{luo2008rapid,pu2011recognizing,cabo2014algorithm}. Cylinder detection methods that rely on the circular shape of poles are only feasible with close-range scans that are mostly available in indoor environments \citep{press2004approaches,aschoff2004algorithms}. For outdoor data, the trunks are actually degenerated to linear shapes, which could be classified using local point classification methods. As for point classification, most previous methods only classify the points into one of three basic categories: linear, planar and volumetric \citep{lalonde2006natural,demantke2011dimensionality}. \cite{yokoyama} introduced the pole-trunk classification, and \cite{jing} further distinguished between points that are part of a wire and the general linear points. The method we describe in this article incorporate all the point classification categories mentioned earlier. In addition, to the best of our knowledge, this is the first work that is able to compute line-of-sight among poles based on LiDAR data.

\section{Generation of Sites and Connections using 3D Computer Vision}\label{cv}

%\subsection{Overview}

\begin{figure}[t]
\centering\includegraphics[width=1.5in]{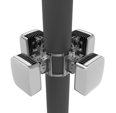}
\caption{A Terragraph site with 4 nodes}
\label{tg}
\end{figure}

Millimeter wave technologies such as Terragraph need to place equipment in {\em distribution nodes} (DN) that connect with each other. These nodes form the backhaul network that enables fast transmission between Fiber POPs and the {\em client nodes} (CN) connected to the (virtual) {\em demand nodes} that represent users.
Figure~\ref{tg} shows a Terragraph site in which multiple antennas (nodes) are deployed in a street pole.

The initial step of the network planning process consists in performing 
a site survey to construct a list of potential sites for DNs. Traditionally, this has involved walking the streets to do a manual inspection of the deployment area with the goal of identifying potential installation locations and estimating RF propagation ranges. 
%Evidently, this was expensive and time consuming, so the approach does not scale to large deployments.
We built a methodology that allowed us to automate this process relying on 3D computer vision and machine learning. Our approach effectively reduces the time required for site surveys, while also also making the process more accurate and comprehensive. 
%For example, a human surveyor will not always be able to accurately determine line-of-sight between the top of two street lights from ground level nor will he/she exhaustively and accurately identify every possible installation location. 
We use LiDAR, short for {\em Light Detection and Ranging}, which is a remote sensing technology that can measure the distance to any surface using lasers.
The output of LiDAR data collection for an area is a {\em point cloud} that consists of a list of points with their three-dimensional $(x, y, z)$ coordinates.
We collected the raw data using vehicle-mounted LiDAR system at average sampling of a point every 3\,cm with a measurement error of up to 4\,cm. Subsequently we post-process the data by geo-referencing it (to a unique world coordinate system). Finally, to allow us to perform computer-vision analysis  efficiently, we down-sample the point cloud to a density of a point every 10\,cm. As an example,
Figure~\ref{pre_process}(a) shows a LiDAR point cloud captured for a small urban area.

%In millimeter wave network design for Terragraph, 
Our goal is to use the LiDAR data as input to identify potential deployment sites. However, each point in the LiDAR point cloud only contains information about its coordinates, and possibly the RGB color values. It is best thought as pixels forming a bitmap image, but in 3D.
This means that we receive an input with no labels or annotations describing what each point represents. To transform the point clouds into data that can be further processed and fed into the network design problem, we apply 3D computer vision algorithms to detect poles (e.g., light poles, electricity poles, traffic lights, etc.) and label them as potential sites where network equipment can be deployed. Furthermore, with LiDAR data capturing all street objects, we analyze whether there is line-of-sight between nearby sites to propose clear paths for radio propagation.
Our LiDAR processing methodology consists of three main modules: pre-processing, pole detection, and line-of-sight analysis, which we describe in detail in the following sections.

\subsection{LiDAR Pre-Processing}

\begin{figure}[h]
\centering\includegraphics[width=6.5in]{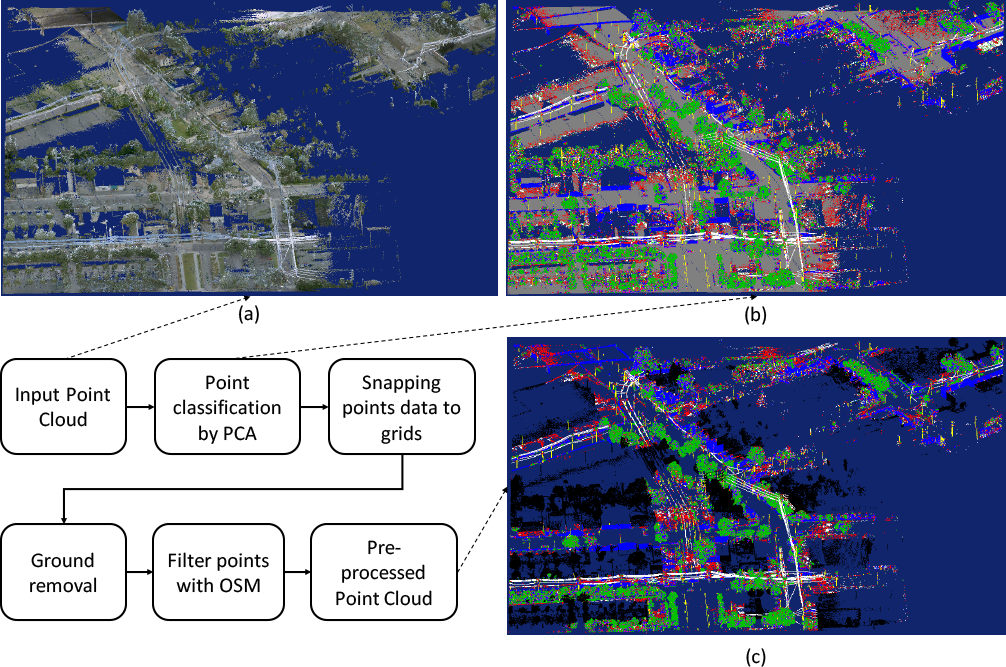}
\caption{Flow diagram of LiDAR point cloud data pre-processing. (a) Example of a LiDAR point cloud for a small urban area; (b) PCA point classification result for the sample LiDAR point cloud (Yellow: Vertical linear shapes; White: Wires; Red: Other linear shapes; Blue: Building facades; Grey: Ground; Green: Tree foilage); (c) Example of ground points removal and OSM filtering (Black: Points filtered by OSM).}
\label{pre_process}
\end{figure}

The first step of the pre-processing module is to perform point classification using {\em Principal Component Analysis} (PCA), followed by hashing the classified and filtered points using UTM-projected $(x,y)$ coordinates into 1m-by-1m grids for efficient neighborhood look-up operations. We end the pre-processing by removing ground points based on the classification results and, optionally, removing irrelevant points. Figure~\ref{pre_process} provides a block diagram for the pre-processing flow with sample data and result at each major stage. In the following sections, we explain those steps in more detail.

\subsubsection{Point Classification by PCA}\label{pca}

We detect poles relying on the fact that they contain a linear vertical stem. In addition, depending on the specific type of pole, there might be other linear components connected to the stem, such as a pole arm or wires. Instead, other street objects are usually characterized by planar structures (e.g., buildings), or by volumetric structures (e.g., trees). Therefore, classifying linear, planar and volumetric structures is a crucial step to classify points in the cloud. 
Inspired by the algorithm developed by \citet{jing}, 
for each point $P$, PCA is done in the neighborhood of radius $r$ centered at $P$, using all points within the radius. The points in the neighborhood can be efficiently searched using a $k$-dimensional tree. 
The three resulting eigenvalues $\lambda_1 \ge \lambda_2 \ge \lambda_3$ produced by PCA satisfy:

\begin{itemize}
\item 
$\lambda_1 \gg \lambda_2 \simeq \lambda_3$ for a linear neighborhood;
\item
$\lambda_1 \simeq \lambda_2 \gg \lambda_3$ for a planar neighborhood; and
\item
$\lambda_1 \simeq \lambda_2 \simeq \lambda_3$ for a volumetric neighborhood.
\end{itemize}

%{\bf [NS: this is not clear, lambda, alpha, beta not defined. Value $d$ defined below is not used. Do we use the $e_i$ above or the $S_i$ below to classify?] - [GP: $\lambda$ is the same as $e$ - changed all $e$ to $\lambda$. Also added explanation about alpha, beta and d in the followed paragraph.]}

We apply the distribution features as in \citet{yokoyama}:
\begin{equation}\label{eqn1}
\begin{cases}
S_1:=\lambda_1-\alpha\lambda_2\\
S_2:=\lambda_2-\lambda_3\\
S_3:=\beta\lambda_3\\
\end{cases}
\end{equation}
where the coefficients $\alpha$ and $\beta$ represent weights set empirically to adjust the sensitivity to each class. 
We use indices in the set $\{1,2,3\}$ to indicate point neighborhoods that are linear, planar or volumetric, respectively. 
The dimensionality feature $d := \arg\max_{i \in \{1,2,3\}} S_i$ is then used to classify the distribution of point $P$'s neighborhood based on the strongest feature $S_i$. 
For a linear neighborhood, the dominant eigenvector $v_1$ (corresponding to $\lambda_1$) indicates the direction of the linear distribution, so we can use \eqref{eqn2} below to classify whether the neighborhood is likely to be a pole stem, which should be vertical. 
To test this condition, we calibrate a threshold $\theta_t$, and use the condition:

\begin{equation}\label{eqn2}
\left| \frac{\vec{v_1}}{||\vec{v_1}||}\cdot(0,0,1) \right|>\theta_t
\end{equation}

Similarly, for a planar neighborhood, we use the direction of the eigenvector $v_3$ (corresponding to the third-largest eigenvalue $\lambda_3$) to determine if it represents the ground. If so,
 $v_3$ should be vertical since it would normal to a horizontal plane. 
 In the above classifications, points which are part of wires will also be classified as linear points. However, it's important to discern wires from other linear neighborhoods because

\begin{itemize}
\item Wires are not part of the pole and should be removed from the detected pole components;
\item Wires usually connects multiple poles and will prevent the clustering of individual poles if not removed;
\item Wires can be used as a feature to recognize utility poles from other structures.
\end{itemize}

We rely on two properties of wires to differentiate them from other linear neighborhoods:

\begin{itemize}
\item Wires are mostly thinner than other linear pole components;
\item The principal direction of wires is typically horizontal.
\end{itemize}

As suggested in \citet{jing}, 
we classify points that are part of a wire using a stricter distribution feature $S'_1$, similar 
to $S_1$ in \eqref{eqn1} but with a much larger value of $\alpha'$:
\begin{equation}\label{eqn3}
\begin{cases}
S'_1:=\lambda_1-\alpha'\lambda_2>S_2,S_3\\[3mm]
\left| \dfrac{\vec{v_1}}{||\vec{v_1}||}\cdot(0,0,1) \right|<\theta_w
\end{cases}
\end{equation}

Based on experimentation with the data we collected, though, we found that \eqref{eqn3} alone is not enough to classify wire points because sometimes several wires lining up in parallel can result in a planar neighborhood.
We use a smaller neighborhood with radius $r'<r$ whenever a planar point is found in radius $r$, and we re-classify it as a wire point if $S'_1$ is the dominant feature in the $r'$-neighborhood.

After calibrating these procedures with actual data, we set $\alpha = 3$, $\beta = 2$, $\alpha' = 20$, $r = 1\,\text{m}$, and $r' = 0.3\,\text{m}$.
Figure~\ref{pre_process}(b) displays an example of the result of the classification where different colors represent different components.

%{\bf [BK: what is the computational complexity of this procedure?  Are you running one PCA for every point in the point cloud?] - [GP: added analysis and explanation in the next paragraph.]}

The computational complexity of these procedures is roughly $O(nr^3)$. Here, $n$ is the number of points in the LiDAR point cloud, which can be very large (i.e., of the order of a billion points). This makes computing PCA for every point very time consuming. We improve the runtime by only considering a sample of points: we select a single point in each neighborhood of radius $R_{pca}$  (which may be different from the radius $r$ used to compute PCA), and the classification of the point is propagated to all other points within the neighborhood. Experiments show that setting $R_{pca} = 0.5\,\text{m}$ results
in a significant speedup with few point mis-classifications that won't lead to a large difference in pole detection results.
%, while significantly speeding-up the process. 
%{\bf [GP: @JH seems in your previous comment you confused $r$ and $R_{pca}$. Please check if the updated description is clear enough.] - [JH: I see, now it looks clear to me. (made a minor change (assigned -> propagated)) Thank you!]}

\subsubsection{Snapping Data in Point Cloud to a Grid}\label{hash}

In a typical point cloud, points are stored in an unordered data structure and can be permuted in any way without changing the data  represented by the cloud. 
Large-scale urban point clouds consist of geo-registered uniquely-identifiable $(x,y)$ coordinates, split in multiple separate segments of data.
Storing data this way, albeit simple, makes it inconvenient and  time-consuming to look up points in a given neighborhood since all the points from all segments of data must be considered to select those at close proximity. 
A better storage mechanism is needed to allow us to obtain points in a certain neighborhood, which is required to
look up external data and municipal databases with geo-coordinates, and check line-of-sight, as described in Section~\ref{sec:los}.

To improve the cloud point retrieval strategy, we hash data into a 1m-by-1m grid. Each {\em grid point}%
\footnote{We highlight the difference between `cloud point' which refers to the original points and `grid point' which refers to the vertices that compose the grid. As we discuss here, each cloud point is mapped to a grid point, and each grid point represents a set of cloud points.}
has two hashed ids using the integer values of UTM-projected $(x,y)$ coordinates. The elevation $z$ is not hashed because it is usually bounded by the building height. 
For each grid point, we store  an unordered list containing all cloud points snapped to it, together with the number of cloud points of each class.
This data structure permits us to efficiently look through all points near specific geo-coordinates, by simply checking the grid points with the hashed ids corresponding to coordinates nearby. Moreover, merging data from multiple segments can be simply done by merging grid points with the same hash ids. We rely on this data structure to execute the procedures described in the following sections efficiently, including filtering, retrieval of pole component points, and line-of-sights analysis.

\subsubsection{OSM Filtering and Ground Removal}\label{sec:osmfilter}

{\em Open Street Map} (OSM) \citep{osm} is a collaborative public project to create a free editable map of the world. Its data contains \textit{nodes} and \textit{ways}, where \textit{nodes} mark specific geographical locations, and \textit{ways} connect multiple \textit{nodes} to represent streets. For visual understanding of street objects using ground-based LiDAR point cloud data, it is advisable to remove any points not close to streets because they have a lower point resolution from ground-based LiDAR capturing, and are usually fragmented and noisy, thus can easily lead to false positives in object detection. 
%In the Terragraph project specifically, poles not close to street are usually privately-owned and should not be considered for the network design, which is another reason to filter points with OSM.

With the point cloud data hashed into a geo-referenced grids, as described earlier, it is simple to label points close to OSM \textit{ways}. The idea is for each \textit{way} in the area, find all grid centers within a certain distance (using hash ids / geo-coordinates) and label them as \textit{close}. After processing all \textit{ways} in the area, any grid center not marked as being \textit{close} will be marked as being \textit{far}, and will be filtered out in later processing. All points stored in a grid will follow the label of that grid center.

%
%
%
%
%

%x
%x
%x

Besides filtering points far from OSM ways, we also remove points classified as ground by PCA because they take a large portion in the point cloud and are not considered to be part of any street object. Figure~\ref{pre_process}(c) shows the result of this process with ground points removed and points far from OSM ways colored in black.

\begin{figure}
\centering\includegraphics[width=6in]{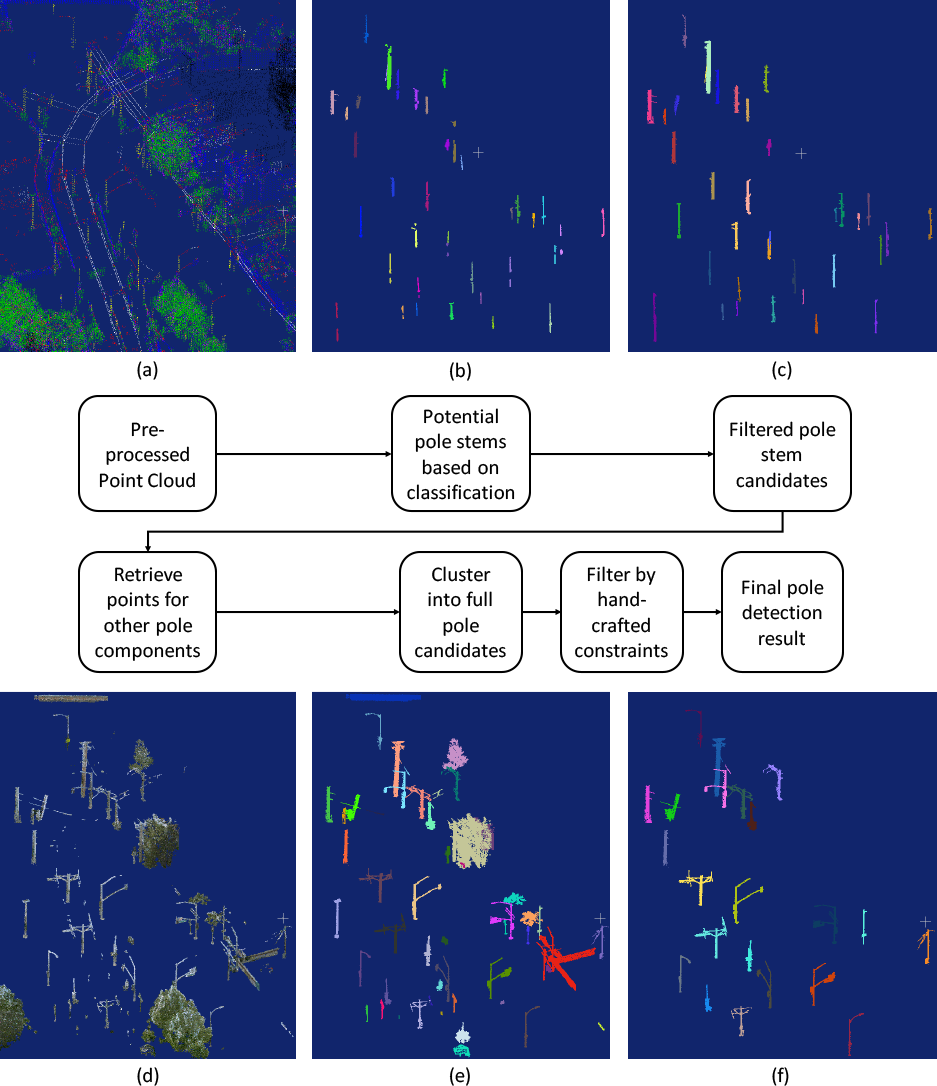}
\caption{Flow diagram of the pole detection process after LiDAR point cloud data is pre-processed. (a) Pre-processed LiDAR point cloud data as input to pole detection; (b) Cluster points with a vertical linear local shape to locate potential pole stems; (c) Filter pole stem candidates according to heights and ground levels; (d) Retrieve points that are close to the candidate pole stems that might belong to other pole components; (e) Cluster points that are close together to get potential full poles; (f) Final pole detection result after filtering with the constraints.}
\label{pole_detect}
\end{figure}

\subsection{Pole Detection}

%{\bf\large [NS: I think we should provide a few more details to explain this clustering or remove some of the equations. This way we're in the middle ground in which we try to provide details but it's hard to understand. Happy to get together briefly to improve this paragraph.]}

With the point cloud pre-processed, we move to the next step which consists of pole detection. Figure~\ref{pole_detect} provides a block diagram for the pole detection flow with sample results at each step. 

This module starts with the pre-processed LiDAR point cloud data (Figure~\ref{pole_detect}(a)). The first step is to locate potential pole stems by clustering points classified as stem points (linear neighborhoods in vertical direction, see Section~\ref{pca}). This will include not only real pole stems, but also other vertical linear structures such as tree trunks and building edges. To obtain pole stem candidates, we cluster stem points that share the same or nearby $(x,y)$ coordinates, as shown in Figure~\ref{pole_detect}(b). The pole stem candidates are then filtered by attributes such as height and ground level. Ground level is obtained from $z$ coordinates of ground points in the same grid cell as the candidate stem or nearby grid cells. As an example, the result of this filtering procedure is shown in Figure~\ref{pole_detect}(c).

Since in some cases one may prefer to mount equipment in the arm attached to the pole, the algorithm will try to retrieve other components of the pole besides the stem. Points close to the pole stem candidates are retrieved and merged together, as shown in Figure~\ref{pole_detect}(d). 
This can be done efficiently using the grid explained in Section~\ref{hash}, by checking all hash ids that are within a certain distance to stem coordinates. Afterwards, we use a clustering algorithm to group together nearby points into standalone objects (Figure~\ref{pole_detect}(e)), which are considered as potential candidates of full pole clusters. 
Finally we classify these clusters to differentiate poles from other objects like trees or buildings. 
This is done using the constraints listed below. To complete the example, Figure~\ref{pole_detect}(f) illustrates the  poles identified after applying the constraints.

%{\bf [BK: Figures 5 and 6 are hard to see.  Can we zoom in or clarify these somehow?] - [GP: Two solutions: 1 - Split figure 5 into two separate figures and figure 6 into three; but this might create too many figures. 2 - Only show a small sub-area in figure 5 and 6; then they won't be in the same area as figure 2-4. Which solution sounds better?] - [JH: I think showing a small sub-area with detail might be better. For large regions in small figures it's quite hard to see the differences.]}

\begin{itemize}
	\setlength\itemsep{0.1em}
	\item Total height is between 5\,m and 30\,m;
	\item Pole stem base $z$ value is close to ground level (referring to nearby ground points);
	\item Maximum ratio of planar and volumetric points are 60\% and 75\%, respectively;
	\item Minimum ratio of linear stem points and all linear points are 10\% and 20\%, respectively;
	\item Among points within the highest 40\% $z$ range, volumetric points are fewer than 50\% (used to filter trees), unless there are enough wire points nearby (to keep utility poles);
	\item Further clustering points only within the lowest 20\% $z$ range, then total area size of all resulting clusters is less than $2\,m^2$.
\end{itemize}

\subsection{Line-of-Sight Analysis}\label{sec:los}

Since millimeter wave technology requires LoS to enable transmission between nodes, determining the visibility between sites is a crucial aspect when building the input to the network design procedure.
Signal propagation paths between network radio sites have been traditionally surveyed by engineers on-site. This manual approach often proves to be both time-consuming and error-prone. In addition, these surveys typically include visiting sites with a bucket truck, which is costly and inconvenient since it disrupts traffic. Lighter-weight approaches such as using OSM could also be used to validate if buildings block the LoS between sites, but we have seen that these methods fall short of providing the needed accuracy, on the one hand because of missing or wrong information, and on the other because the database does not have information related to other obstacles such as foliage. 

\begin{figure}[h]
\centering\includegraphics[width=6.5in]{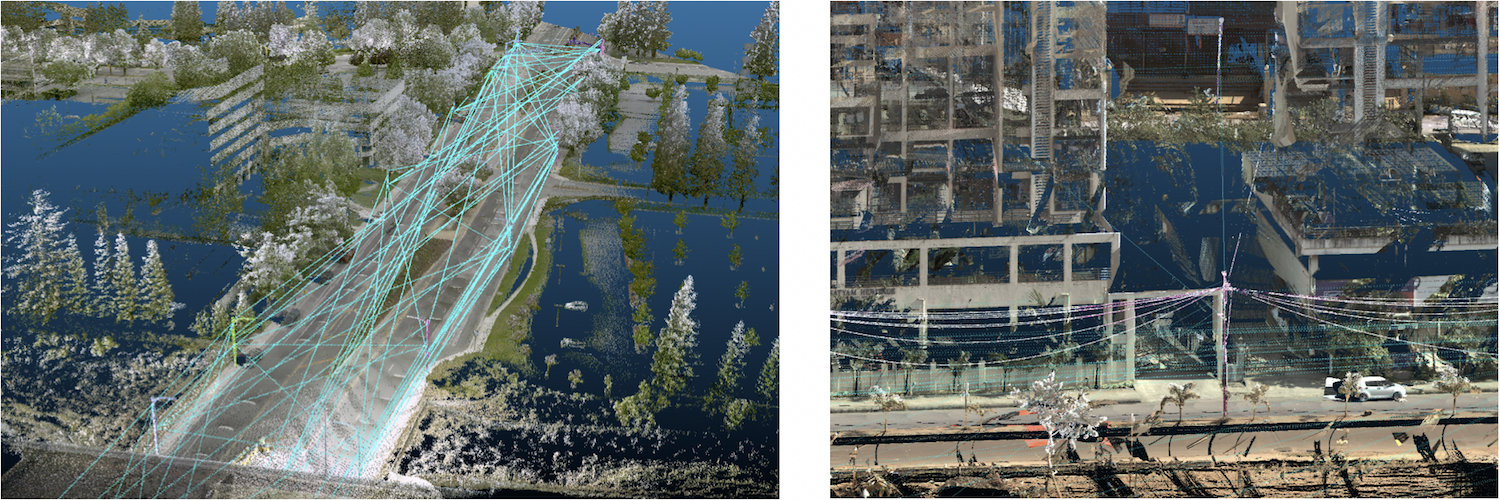}
\caption{Example of automatic line-of-sight analysis results}
\label{line_of_sight}
\end{figure}

Because of those limitations, we turn to LiDAR data because it offers a more accurate 3D representation of all the street objects, and thus it is the ideal medium to perform LoS analysis and obtain available signal paths.
The LiDAR-based LoS analysis algorithm consists of the following steps.
For each pair of sites (i.e., poles detected in the previous step) that are within the maximum transmission distance (e.g., 300 meters), the algorithm checks the number of points close to the 3D segment that connects the coordinates of the two sites at a specific height.
We add the pair of sites to the list of paths with LoS when the number of cloud points close to the segment between those sites does not exceed a threshold.
Counting the points can be done efficiently through the hashed data grid described in Section~\ref{hash} because only the corresponding grid points need to be checked by computing the distance between the coordinates of the original point and the 3D segment.
The LoS algorithm is calibrated by adjusting the height and the threshold point count to maximize precision and recall.
Figure~\ref{line_of_sight} shows the result of the automatic LoS analysis.
%We find pointing angles to account for all possible paths; e.g., going on top of short trees or through building gaps.
The algorithm considers different possible heights to account for all possible signal paths. This allows us to detect the possibility of going on top of short trees or between building gaps. Manually detecting LoS in some of those cases 
is hard if the human surveyor is standing at ground level and estimating the visual reach from the top of a pole. This alone was a frequent cause of mistakes in the manual datasets previously used as inputs.

Alternatively, OSM can be used to propose all LoS with no building occlusion, and subsequently LoS is refined through LiDAR to detect other street obstacles such as trees. This approach is faster than using LiDAR alone, and is specially useful in areas where LiDAR data is incomplete. At the same time, it might overestimate building occlusions in some areas of the world because OSM does not have detailed 3D data of buildings.

\section{Integer Optimization for Terragraph Network Design}\label{ip}

Given the list of potential sites (i.e., poles) and connections (i.e., paths) computed by the methodology described in Section~\ref{cv}, the next step is to select a network spanned by a subset of the potential sites.
To design the network, we need additional data such as the location of Fiber POPs and demand measures as the connectivity requirement. We add POP information as a yes/no attribute to the potential sites.
Demand is represented by demand points. Depending on the planned design, this can be encoded by having demand points representing individual buildings for which there must be connectivity, or by creating a uniform grid (say, by placing a demand point every 30 meters) when one is designing for blanket coverage WiFi provisioning. 
%{\bf [TD: Do we want to refer to other radio planning tools outside this set that provide the input to the WiFi network]}

Since antennas have a limited operation angle and finite throughput, each deployed site (DN) will commonly include multiple antennas pointing in different directions (see Figure~\ref{tg}). We refer to each of these antennas as a {\em sector}. In addition, in our network representation, we also include other connection capabilities of DNs such as WiFi, LTE access points and Internet Fiber POPs as additional (virtual) sectors. In summary, we associate multiple potential sectors to each potential DN, forming a hierarchical network structure. Obviously, we can and will only deploy the sectors if the DN is deployed.

We model the Terragraph problem as a hierarchical network design problem
combined with a flow problem on a network $N(V,A)$ where the nodes $V$ represent the sets of DNs, CNs, POPs, and the demand locations. Each node $v\in V$ is associated to the sectors that are available in it, and network links are included for any two sectors that can connect to each other as explained in Section~\ref{sec:los}. Notice that there are different types of links: two DN sectors or a DN and a CN sector can communicate using Terragraph hardware, a CN sector can be connected to a demand node in that location with a wired Ethernet link, and a WiFi sector can connect to demand nodes that are within transmission range.  
We formulate the problem as an {\em Integer Program} (IP). This methodology gives us the flexibility to express the requirements that the network must adhere to. At a high level, we build the network that satisfy demand at minimum cost. 
We now describe the mathematical programming formulation in detail and, Section~\ref{steiner} explains the heuristics employed to scale the solution procedure and solve real-world problem instances.

\subsubsection*{Inputs to Optimization Model}
The description on the problem relies on the following input parameters for the area where the deployment is being considered.
Besides the DNs and LoS which are computed using the methodology described in Section~\ref{cv}, the rest of the basic input data is gathered manually and processed to generate the following information. The node data is given by:

\begin{description}
\item[$DN$:] Set of potential distribution node sites (Section~\ref{cv}).
\item[$CN$:] Set of potential client node locations (only in configurations where the Terragraph backbone is used to connect buildings to the leaves of the network using wired connections).
\item[$POP$:] Set of potential Fiber POP locations.
\item[$DEM$:] Set of demand locations (depending on the input configuration, this set may include the buildings that need to be connected, one point in the centroid of each city block, or a uniform grid covering the deployment area).
\item[$INT$:] Singleton set containing the source node used in the network flow problem.
\item[$V$:] Set of all sites, defined as $CN\cup DN \cup POP \cup DEM \cup INT$.
\item[$C_i$:] Cost of deployment for location $i \in V$.
\item[$D_i$:] Demand at demand location $i\in DEM$.
\end{description}

\bigskip\noindent
Potential connections are given by:

\begin{description}
\item[$A_{DN}$:] Set of all possible DN-DN connections (as computed in Section~\ref{cv}).
\item[$A$:] Set of all possible connections between nodes in $V$ ($A_{DN}$ is augmented with connections between other node types).
\item[$T_{ij}$:] Connection throughput between nodes $(i,j)\in A$ (maximum transmission capacity if corresponding sectors are used exclusively for this connection).
\item[$Q_{ij}$:] Connection quality between nodes $(i,j)\in A$ (probability that connection works).
\end{description}

\bigskip\noindent
The input related to potential sectors include:
\begin{description}
\item[$\delta_i$:]  Set of potential sectors that can be built on site $i\in V$.
For DN-DN and DN-CN connections, the set $\delta_i$ can be thought as a set of possible orientation angles of the sector when mounted on the street furniture. In addition, sectors may include other connections such as WiFi and LTE access points.
\item[$S$:] Set of potential sectors, defined as $\cup_{i\in V} \delta_i$.
\item[$\Gamma_{is}$:] Set of sites that sector $s\in \delta_i$ at site $i \in V$ can communicate with. Notice that $(i,j) \in A$ only if $j\in \Gamma_{is}$ for some $s \in \delta_i$.
\item[$K_{is}$:] Cost of deploying sector $s \in \delta_i$ at site $i\in V$.
\end{description}

\subsubsection*{Decision Variables in Optimization Model}

The decisions variables used by the IP are given by:

\begin{description}
\item[$z_i$:] is 1 if site $i\in V$ is deployed, $0$ otherwise.
\item[$s_{ia}$:] is 1 if sector $a\in S$ on site $i\in V$ is deployed, $0$ otherwise.
\item[$f_{ij}$:] flow sent through connection $(i,j)\in A$.
\item[$m_{ij}^k$:] flow sent through connection $(i,j)\in A$ for demand location $k \in \text{DEM}$.
\item[$p_{ij}$:] time division multiplexing at connection $(i,j)$.
\item[$r_i$:] is 1 if the coloring of site $i\in DN\cup POP$ is red, 0 otherwise (optional). Colors red and blue represent synchronization periods which may be used by Terragraph to reduce interference between simultaneous transmissions. 
\item[$b_i$:] is 1 if the coloring of site $i\in DN\cup POP$ is blue, 0 otherwise (optional).
\end{description}

\subsubsection*{Formulation of the Optimization Model}

Both inputs and variables are combined in the following formulation.
A crucial objective is to minimize the total cost of ownership, which includes cost of hardware and operational expenses (e.g., remediation costs necessary to make all the selected sites a viable antenna location, rental cost, energy cost, etc). Note that alternative objective functions can easily be considered in this setting (see Section~\ref{maxdemand}). Letting $C_i$ to be the cost of activating site $i\in V$ and $K_{is}$ to be the cost of activating sector $s\in \delta_i$ at site $i\in V$, we would like to find the subset of sites and sectors that provides the minimum total cost:
\begin{align}
\min \quad &\sum_{i \in V} C_i z_i + \sum_{i \in V}\sum_{j \in \delta_i} K_{ij} s_{ij}. \label{obj_mc}	
\end{align}

As mentioned earlier, we introduced a super-source node in $V$ referred to as $INT$ to represent Internet traffic that goes into the network through the POPs. This node includes connections in $A$ to all POPs. Variables 
$f_{ij}\in \mathbb{R}_{\ge 0}$ denote the flow sent through connection $(i,j)$. We ensure that these flows represent valid traffic using flow balance constraints, which state that the total incoming flow is equal to the  sum of the total outgoing flow and the demand at that location:
\begin{align}
&\sum_{j\in V: (j,i)\in A} f_{ji} = \sum_{j\in V: (i,j)\in A} f_{ij} + \tilde{d}_i \quad i \in V \label{flowbalance} 
\end{align}
where $$\bar{d}_i = \begin{cases} &D_i, \quad i\in DEM\\ & -M, \quad i \in INT\\ &0\quad \text{otherwise} \end{cases}$$ and $M = \sum_{k \in DEM} D_k$. As a result of the flow balance constraints, each demand location $i\in DEM$ has at least one path to an active POP that forwards the traffic directed to it.

The binary decision variable $z_i\in \{0,1\}$ represents if site $i$ is selected to be a part of the final network design. A site
may only have nonzero incoming or outgoing flow when it is active:
\begin{align}
&\sum_{j\in V} f_{ij} \leq M z_i, \quad i \in V \label{useloc_fwd}\\
&\sum_{j\in V} f_{ji} \leq M z_i, \quad i \in V \label{useloc_rev}.
%& z_i \in \{0, 1\}, \quad i \in V
\end{align}
Since in this formulation, we satisfy demand at minimum cost, we assume that demand locations are always active:
$z_k = 1$ for $k\in DEM$.

Next, sectors must be selected to enable connections between sites. Indeed, a nonzero flow $f_{ij}$ from site $i$ to site $j$ can only be achieved if the corresponding sector $a$ at site $i$ is deployed.
In addition, a sector at a potential DN or POP site $i$ can be activated only if site $i$ is also active.
\begin{align}
&\sum_{j\in \Gamma_{ia}} f_{ij} \leq M s_{ia}, \quad i \in V, a \in \delta_i \label{usesec_fwd}\\
&\sum_{j\in \Gamma_{ia}} f_{ji} \leq M s_{ia}, \quad i \in V, a \in \delta_i \label{usesec_rev}	\\
&s_{ia} \leq z_i, \quad i \in DN\cup POP, a \in \delta_{i} \label{sectorlocation}.
%&s_{ia} \in \{0,1\}, \quad i\in V, a \in \delta_i.
\end{align}

The next set of constraints encode the requirements for CN connectivity coming from the Terragraph design.
At potential CN locations that are active, we assume that there is at most one sector connecting to a demand location and at most one sector connecting to another DN or POP. The following constraints ensure that in an active CN $i$, the number of sectors facing a DN/POP can be at most one, and that the number of sectors facing a demand location can be at most one.
\begin{align}
&\sum_{a \in \delta_i: |(DN\cup POP) \cap \Gamma_{ia}| \geq 1} s_{ia} \leq z_i, \quad i \in \text{CN} \label{cn1}\\
&\sum_{a \in \delta_i, |DEM \cap \Gamma_{ia}| \geq 1} s_{ia} \leq z_i, \quad i \in \text{CN} \label{cn2}.
\end{align}
We also ensure that each demand location $k$ is covered by exactly one CN location. This is typically used to select the best CN among alternatives to connect to a building. 
\begin{align}
&\sum_{i\in CN: (i,k) \in A} z_i \leq 1, \quad k\in DEM \label{onecnperdem}	
\end{align}

If a point-to-multipoint configuration is used, Terragraph sectors send and receive transmissions using {\em time division multiplexing} (TDM). We model TDM using decision variables $p_{ij}\in \mathbb{R}_{\ge 0}$ that represent the fraction of
time $i$ uses to transmit to $j$. We assume that a connection $(i,j)\in A$ can be done using a unique pair of sectors in $i$ and $j$. 
\begin{align}
&f_{ij} \leq p_{ij} T_{ij}, \quad (i,j)\in A \label{eff_cap} \\
&\sum_{j\in \Gamma_{ia}}  p_{ij} \leq s_{ia}, \quad i \in V, a \in \delta_i \label{tdm1}\\
&\sum_{j\in \Gamma_{ia}}  p_{ji} \leq s_{ia}, \quad i \in V, a \in \delta_i \label{tdm2}.
%&p_{ij} \geq 0, \quad (i,j)\in A.
\end{align}
Constraint~\eqref{eff_cap} states that the flow on connection $(i,j)$ cannot exceed its effective throughput. For a sector $a$ at site $i$, \eqref{tdm1} and \eqref{tdm2} state that the TDM fractions cannot exceed 100\%. 
Both constraints can be merged into one if transmission and reception cannot be done at the same time.
If sector $a$ is not active, these constraints prevent transmissions between $i$ and $j$. 

Finally, sites may have polarities, which are used to decrease the interference caused by other signals over the area. Ideally, the designed network should form a bipartite graph and sites should belong to two distinct classes without transmissions among sites belonging to the same class. The two polarities are  used by The Terragraph protocol synchronously transmits from sites with one polarity in one period and then from sites with the other polarity in the next one.
This is an easy version of a well known problem in graph theory called the {\em graph coloring problem}. 
The following set of constraints assign a polarity to each site such that no adjacent sites $i$ and $j$ have the same polarity. 
We let $r_i$ and $b_i$ be the binary decision variables representing the two possible colors of a site $i\in DN \cup POP$. When the site is deployed, exactly one of those variables is set to 1 to designate the corresponding color.
\begin{align}
&r_i + b_i = z_i, \quad i \in \text{DN}\cup\text{POP} \label{bipartite1}\\
&p_{ij} \leq r_i + r_j, \quad (i,j) \in A,~i, j \in \text{DN}\cup\text{POP}  \label{bipartite2}\\
&p_{ij} \leq b_i + b_j, \quad (i,j) \in A,~i, j \in \text{DN}\cup\text{POP} \label{bipartite3}.
%&r_i \in \{0,1\}, \quad i\in \text{DN}\cup\text{POP}\\
%&b_i \in \{0,1\}, \quad i\in \text{DN}\cup\text{POP}
\end{align}
%Constraint \eqref{bipartite1} ensures that an active site can either be blue or red but not both and if the site is not active, then there is no coloring assignment. 
Constraints \eqref{bipartite2} and \eqref{bipartite3} ensure that two sites that have direct communication have different color assignments.

Finally, we need to provision the network with service quality exceeding the minimum requirements. 
We approximate the probability of all demand locations can be successfully connected by taking the weighted average of the log likelihood and constrain this to be above a threshold. Furthermore, we limit the average latency (as represented by the number of hops between POPs and demand locations).
\begin{align}
&\sum_{i, j\in CN\cup DN\cup POP} \log (Q_{ij})  (f_{ij}/M) \geq \log (1-\alpha)   \label{quality}\\
&\sum_{i, j\in CN\cup DN\cup POP} (f_{ij}/M) \leq H, \label{hops}
\end{align} 
where  $1-\alpha$ is a threshold for the minimum probability of providing connectivity to all demand locations, and $H$ is the maximum number of hops that a signal can travel. Both of these constants are tuning parameters that are set to ensure that the quality of service of the resulting solution is appropriate for the deployment that is being planned.

\subsection{Alternative Formulation: Maximize Connectivity for a Budget}\label{maxdemand}
The formulation is flexible enough to consider the case where there is a fixed budget $B$ for the deployment, and the network is built to maximize the connection coverage.
There are several ways to model this approach. We choose to minimize the total units of unsatisfied demand to avoid additional binary variables. Let $y_k\in [0,D_k]\subset \mathbb{R}$ be the amount of unsatisfied demand at location $k\in DEM$. The resulting objective function \eqref{obj_mc} is
\begin{align}
\min \quad \sum_{k \in \text{DEM}} y_k.
\end{align}
The values of $y_k$ are defined by a modified flow balance constraint for demand nodes:
\begin{align}
&\sum_{j\in V} f_{jk} - \sum_{j\in V} f_{kj} = D_k - y_k, \quad k\in DEM \label{servfbcont}.
%&0\leq y_k \leq D_k, \quad k\in \text{DEM}. \label{servfbcont2}
\end{align} 
Finally, we add the budget constraint
\begin{align}
\sum_{i \in V} C_i z_i + \sum_{i \in V}\sum_{a \in \delta_i} K_{ia} s_{ia} \leq B.
\end{align}

\subsection{Solving the Problem}\label{steiner}

We solve the IP described earlier with a commercial solver, which can find feasible solutions and bounds to the optimal solution given by the linear programming relaxation. The optimal solution is found by intelligently searching the combinations of binary values that are feasible. Since ultimately the network design problem is NP~hard, the solution time will degrade as instances get larger. 
We incorporated heuristic algorithms to quickly generate feasible solutions to speed up the branching procedure done when enumerating feasible solutions.

%The integer programming approach above is quite powerful. As the project evolves, we can add new constraints, change the objective function easily. However, the combinatorial nature of the problem causes the solution times to increase exponentially with increasing problem sizes. As a result, alternative heuristic algorithms with fast solution times are necessary.

To simplify the problem, we solve the  Terragraph network design problem into two stages: we first select the active sites and, then, we find sectors in those sites that allow us to deploy that network topology. We designed a  heuristic for the first stage based on the similarity of this problem to the classical \emph{node weighted Steiner tree problem} \citep{karp}, defined as follows.
For a graph $G=(V,A)$ and a subset of \emph{terminal nodes} $S\subseteq V$, a {\em Steiner tree} is a subgraph of $G$ that spans all the terminal nodes. The node weighted Steiner tree problem aims to find a Steiner tree of minimum total node weight. Although, this problem is known to be NP-hard, it has been deeply studied and there are many heuristics proposed in the literature that perform well.
See \cite{winter1987steiner} for a survey on Steiner problems in networks.

The goal of the Steiner tree problem is to connect the super-source node $INT$ and all the demand locations in $DEM$. The node weights are set to the cost of activating a site. As a result, the node weighted minimal Steiner Tree is equivalent to a sub-graph of the input network $G$ that connects the super-source to all demand sites with minimal total site activation cost. In this study, we use a heuristic algorithm that makes use of shortest paths and minimum spanning trees to find a near optimal solution which is outlined in Algorithm~1 of \cite{sadeghi2013steiner}. The algorithm starts with a graph $G^{\prime} = (\{INT\}, \emptyset)$. For our shortest path and minimum spanning tree calculations, we assign the edge costs as the activation cost of the site that the edge is originated from. At each iteration, we find a demand vertex that is not in $G^{\prime}$ with the shortest distance to any of the vertices in $G^{\prime}$. Then, we expand $G^{\prime}$ with this demand vertex and the shortest path connecting it to $G^{\prime}$. Once all the demand vertices are in $G^{\prime}$, we obtain the minimum spanning tree $T$ induced from the vertices and edges of $G^{\prime}$. As a last step, we remove the non-terminal vertices in $T$ with degree one since they are not necessary for connecting the terminal nodes. Consequently, the resulting tree $T$ is a Steiner Tree that connects the super-source vertex to the demand locations.

\begin{figure}[t]
\centerline{
\includegraphics[width=5in]{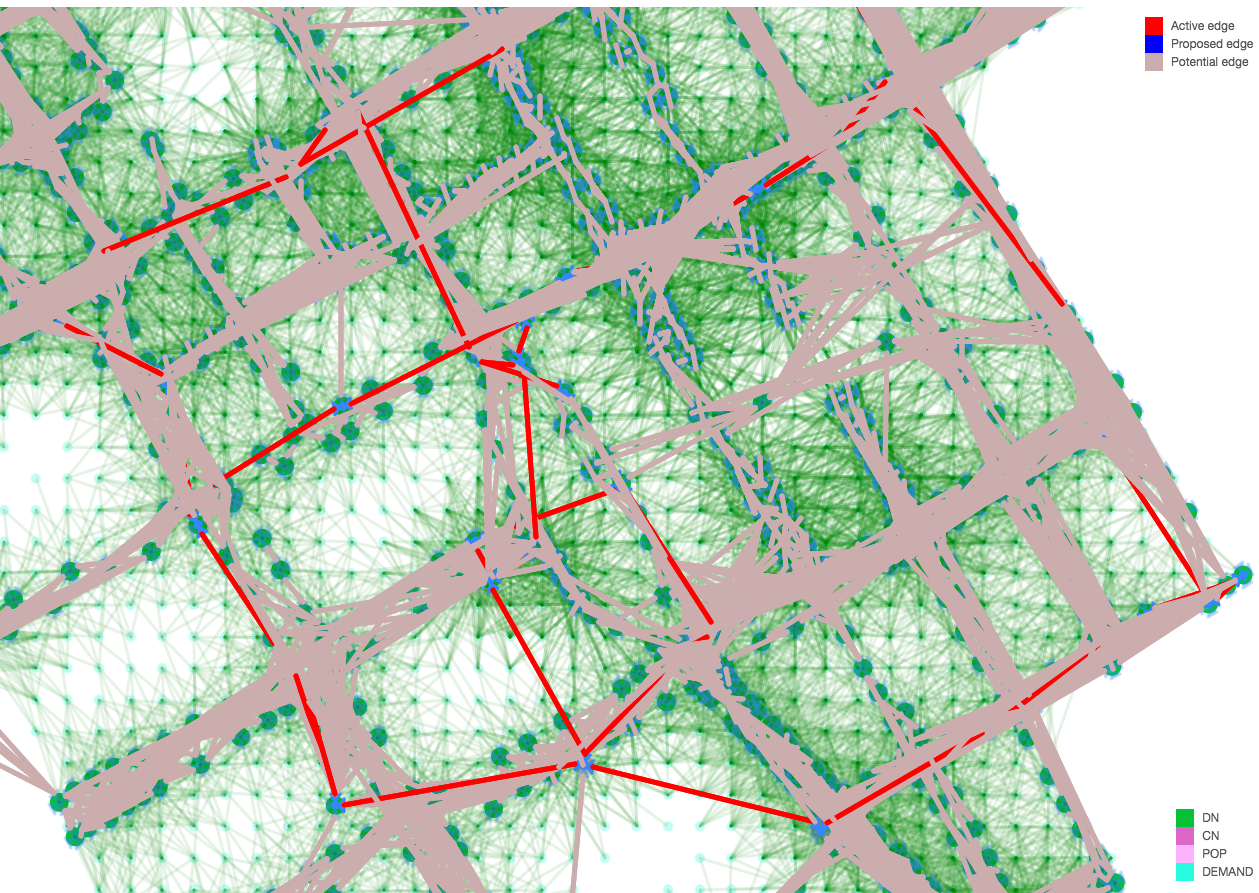}
}
\caption{Input to the network problem instance consisting of sites, connections and demands.}
\label{fig:input}
\end{figure}
\begin{figure}[t]
\centerline{
\includegraphics[width=5in]{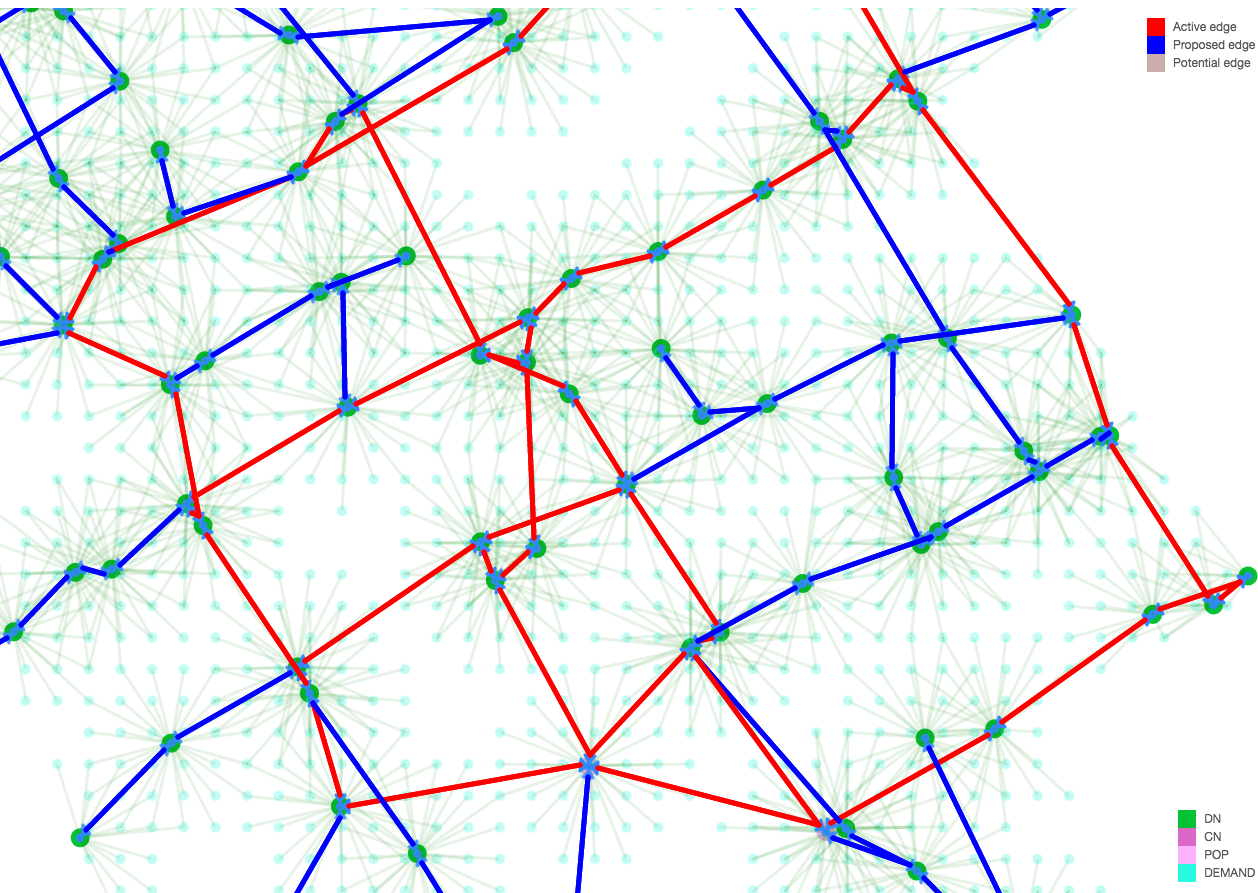}
}
\caption{Output network design computed from the input shown in Figure~\ref{fig:input}.}
\label{fig:output}
\end{figure}

\begin{figure}
\begin{center}
%\includegraphics[width=3.1in,trim={75mm 110mm 212mm  0},clip]{SJCArea2ExtendedPotential.png}$~$%\\[5mm]
%ncludegraphics[width=3.1in,trim={75mm 110mm 212mm 0},clip]{SJCArea2ExtendedProposed.png}
\includegraphics[width=4.1in,trim={75mm 110mm 212mm  0},clip]{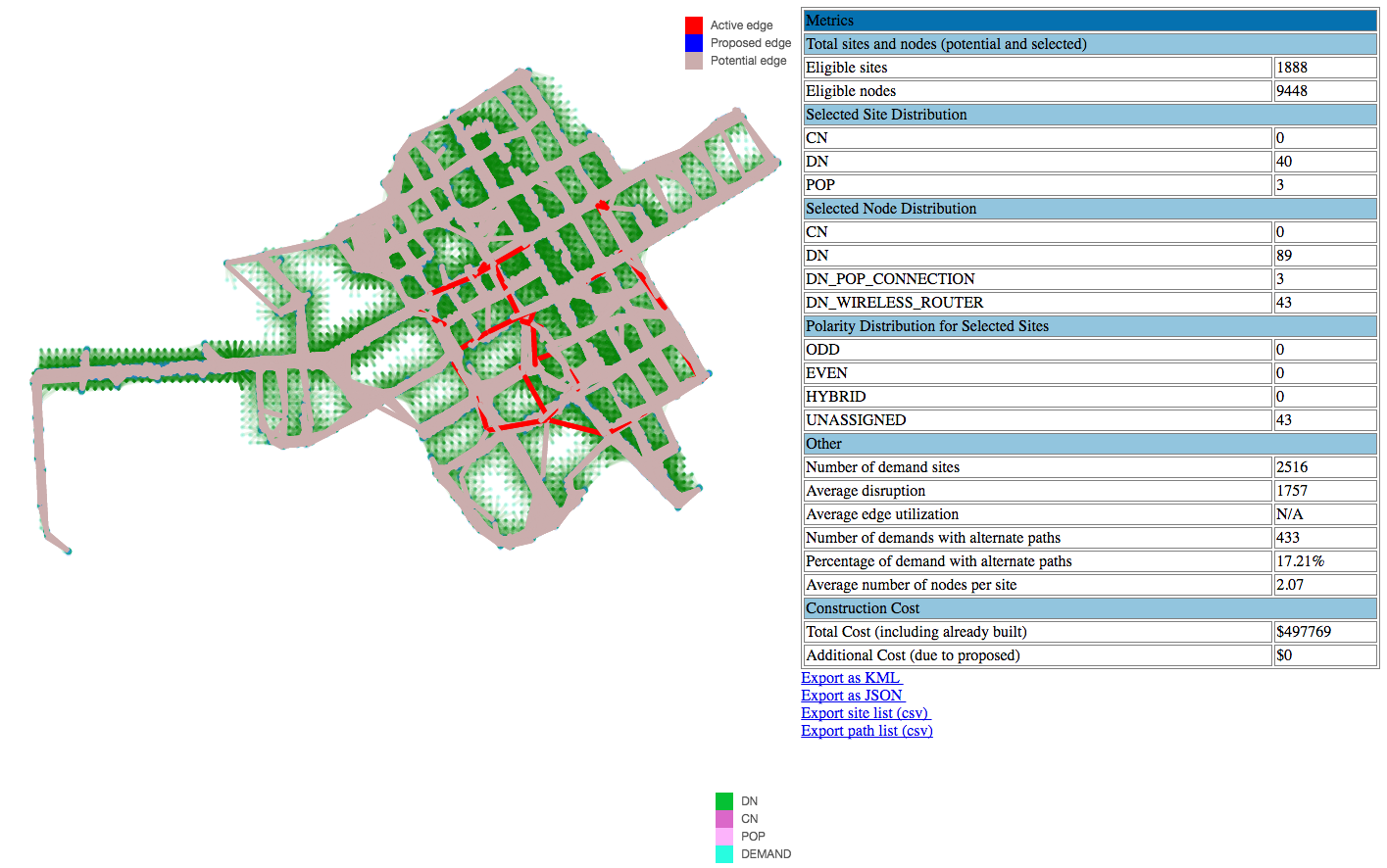}\\[5mm]
\includegraphics[width=4.1in,trim={75mm 110mm 212mm 0},clip]{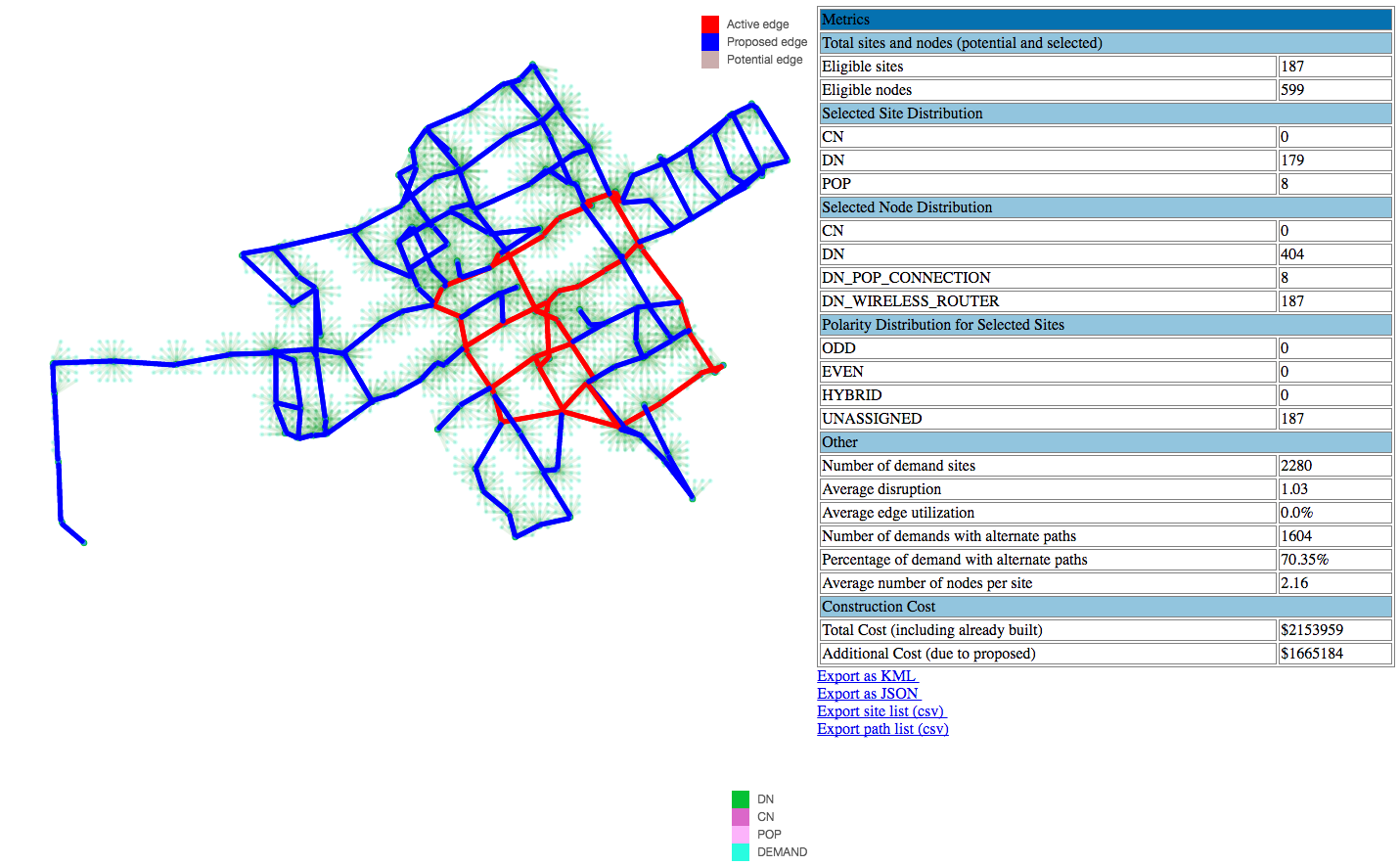}
\end{center}
\caption{Enlarged view of the input and output corresponding to problem instance in Figures~\ref{fig:input} and~\ref{fig:output}.}
\label{fig:output0}
\end{figure}
 \begin{table}[p]
 \caption{Some metrics computed by the visualization module}\label{outputstats}
$~$\\
\centering
\begin{tabular}{lrr}
    \toprule
    \toprule
metric & possible & selected\\\midrule
DN sites & 1888 & 187 \\
Antenna nodes & 9448 & 599 \\
WiFi access points & & 187 \\
Demand vertices & 2516 & 2280 \\
Demand vertices with alternate paths &  & 1604 \\
\% demand vertices with alternate paths &  & 70\%\\
Average nodes per site & 2.07 & 2.16\\
    \bottomrule
    \bottomrule
\end{tabular}
\end{table}

After a solution to the first stage is found, we set variables $z_i$ to 1 for the sites included in the Steiner tree and solve the IP described earlier to complete the solution. Fixing the Steiner tree substantially reduces the complexity of the problem and, consequently, we can solve larger instances much faster. 
% Alternatively, we can solve a secondary Steiner tree heuristic for the second stage as well. Once the first stage decisions are given, let nodes represent sectors that can be built on the locations selected in first stage.  

Figures~\ref{fig:input}-\ref{fig:output0} provide an example corresponding to a test instance. The first figure displays the potential sites (large green vertices) and connections (grey edges) as computed by the Computer Vision algorithm explained earlier. In addition, we include the grid that represents demand (small light blue vertices) and connections between demands and sites (green edges). The example considers a previously-installed backbone that provides the initial connectivity (red edges) that must be augmented to connect as many of the demand vertices as possible at minimum cost, satisfying the constraints specified by the formulation. After running the network design algorithm, we obtain a solution, which we represent in Figure~\ref{fig:output}. It shows the actual sites (large green vertices) and connections (blue edges). As required, the network was completed to connect as many demand nodes (small light blue vertices) as possible at minimum cost. For a given solution, we have built post-processing software that allows an operator to display the networks shown in the figures and modify the solution manually if desired. In addition, we compute key metrics that can be used to compare across possible alternative designs. The metrics for the example are shown in Table~\ref{outputstats}. Other possible metrics that one may compute include throughput, utilization, latency, reliability, availability, and cost.

\begin{table}[p]
  \caption{Variation of network design output metrics for different specifications of throughput requirement, which is shown in the demand column expressed as multiples of the basic throughput requirement.}
  \label{tab:sensitivity}
  $~$\\
  \centering
  \begin{tabular}{ccccccc} 
\toprule\toprule
\multicolumn{1}{c}{Demand} & \multicolumn{1}{c}{Solution} & \multicolumn{1}{c}{Optimality}  & \multicolumn{1}{c}{Antenna}  & \multicolumn{1}{c}{WiFi} & \multicolumn{1}{c}{DN} & \multicolumn{1}{c}{Fiber} \\
%	 \cmidrule(l){7-8}
    \multicolumn{1}{c}{(mult.\ of base)} & \multicolumn{1}{c}{time (s)}  & \multicolumn{1}{c}{gap} & \multicolumn{1}{c}{nodes} & \multicolumn{1}{c}{access points} & \multicolumn{1}{c}{sites} & \multicolumn{1}{c}{POPs} \\
    \midrule
    1     & 180.9 &  8\%   & 149        & 78    & 86    & 1 \\
    2     & 137.5 &  9\%   & 146        & 76    & 85    & 1 \\
 %   3     & 317.5 &  9\%   & 146        & 72    & 83    & 3 \\
    4     & 228.2 &  7\%   & 147        & 75    & 82    & 2 \\
  %  5     & 261.1 &  11\%  & 148        & 76    & 85    & 3 \\
    6     & 142.1 &  7\%   & 150        & 78    & 83    & 3 \\
   % 7     & 98.4  &  1\%   & 155        & 80    & 83    & 3 \\
    8     & 161.8 &  9\%   & 146        & 78    & 84    & 3 \\
   % 9     & 228.0 &  9\%   & 159        & 79    & 84    & 4 \\
    10    & 187.1 &  7\%   & 148        & 78    & 84    & 4 \\
   % 11    & 196.9 &  7\%   & 149        & 78    & 83    & 5 \\
    12    & 222.9 &  8\%   & 152        & 76    & 85    & 4 \\
   % 13    & 121.8 &  6\%   & 161        & 77    & 82    & 5 \\
    14    & 132.5 &  8\%   & 154        & 76    & 85    & 6 \\
   % 15    & 190.6 &  11\%  & 153        & 80    & 87    & 5 \\
    16    & 131.3 &  0\%   & 167        & 74    & 86    & 6 \\
   % 17    & 149.6 &  9\%   & 168        & 81    & 89    & 6 \\
    18    & 124.0 &  3\%   & 161        & 73    & 86    & 6 \\
%    19-25 & -     & -     & -          & -     & -     & - \\
    \bottomrule
    \bottomrule
    \end{tabular}
\end{table}

% We ran the two stage approach where in the first stage we decided which sites to activate and in the second stage, which nodes to activate on these selected sites. The optimization on each stage was ran with a 30 minute time limit. Moreover, we stopped the optimization when a feasible solution with a total cost within at most 15% of the optimal cost value.
 
Finally, we perform a sensibility analysis to illustrate the benefits of having an automatic network design framework. 
We ran our algorithm for different throughput requirement for each demand vertex on a subset of the previous instance with 1171 demand vertices and 182 potential DN sites. 
The demand is measured in multiples of a base level assigned for the regular instance. The runs were stopped when we either reached an optimality gap of under 10\% or when the solution time reached 30 mins.
In Table~\ref{tab:sensitivity}, we observe how the number of active DN sites, fiber POPs, antenna nodes, and WiFi access points change with the throughput requirement. DN sites are not sensitive to throughput requirements in the range we considered, but additional nodes and fiber POPs become necessary when throughput requirements are high. 
We reach a solution within 3 minutes on average and increasing throughput requirements does not affect solution times substantially.

\section{Conclusions}\label{concl}
We have presented a fully automated network planning system that enables rapid deployment of millimeter-wave backhaul and access networks. 
This system has enabled us to design millimeter wave networks like Terragraph in dense urban areas more efficiently, faster and with more accuracy than manual methods.
We use 3D computer vision to analyze 3D city LiDAR data and automate detection of sites, line-of-sight and signal propagation information. Then we apply network optimization to compute deployments that jointly minimize cost and maximize demand served, taking into account the technological constraints that make a deployment feasible. Previously, it had been common to spend weeks on manual data collection and processing, then iterate with more manual work as requirements changed or some sites became infeasible for technical reasons. With the proposed approach, we collect 3D LiDAR data once, and then all iterations are done automatically using 3D computer vision and network optimization, with large savings in time and cost.

In the next phase of our work, we plan to evolve this planning system into an open, collaborative platform that can be used to design dense urban networks, which we believe will serve as a catalyst for widespread adoption of millimeter-wave technology.
%\clearpage

\bibliographystyle{chicago}
\bibliography{biblio}

\end{document}